\documentclass[referee]{aa}
\usepackage{amssymb}
\usepackage{graphicx}
\usepackage{txfonts}
\begin{document}

 \title{Bethe-Heitler cascades as a plausible origin of hard spectra in distant TeV blazars}


   \author{Y.G. Zheng
          \inst{1}
          \and
            C.Y. Yang\inst{2,3}
          \and S.J. Kang\inst{4}
          }

   \offprints{Y.G. Zheng}
   \institute{Department of Physics, Yunnan Normal University, Kunming, 650092, China\\
              \email{ynzyg@yun.edu.cn}
         \and
            Yunnan Observatories, Chinese Academy of Sciences, Kunming 650011, China
         \and
            Key Laboratory for the Structure and Evolution of Celestial Objects, Chinese Academy of Sciences
         \and
           Department of Physics and Electronics Science, Liupanshui Normal University, Liupanshui, Guizhou, 553004, China
             }

   \date{Received ; accepted }


  \abstract
   {Very high-energy (VHE) $\gamma$-ray measurements of distant TeV blazars can be nicely explained by  TeV spectra induced by ultra high-energy cosmic rays.}
   {We develop a model for a plausible origin of hard spectra in distant TeV blazars.}
   {In the model, the TeV emission in distant TeV blazars is dominated by two mixed  components.  The first is the internal component with the photon energy around 1 TeV  produced by inverse Compton scattering of the relativistic electrons on the synchrotron photons (SSC) with a correction for extragalactic background light absorbtion and the other is the external component with the photon energy more than 1 TeV  produced by the cascade emission from high-energy protons propagating through intergalactic space.}
   {Assuming  suitable model parameters, we apply the model to observed spectra of distant TeV blazars of 1ES 0229+200. Our results show that 1) the observed spectrum properties of 1ES 0229+200, especially the TeV $\gamma$-ray tail of the observed spectra, could be reproduced in our model and 2) an expected TeV $\gamma$-ray spectrum with photon energy $>$1 TeV of 1ES 0229+200 should be comparable with the 50-hour sensitivity goal of the Cherenkov Telescope Array (CTA) and the differential sensitivity curve for the one-year observation with the Large High Altitude Air Shower Observatory (LHAASO).}
   {We argue that  strong evidence for the Bethe-Heitler cascades along the line of sight as a plausible origin of hard spectra in distant TeV blazars could be obtained from VHE observations with CTA, LHAASO, HAWC, and HiSCORE.}

   \keywords{BL Lacertae objects: individual(1ES 0229+200)--radiation mechanisms:non-thermal}

   \maketitle
%
%

\section{Introduction}

A blazar is a special class of active galactic nucleus (AGN) with a non-thermal continuum emission that arises from the jet emission taking place in an AGN whose jet axis is closely aligned with the observer's line of  sight (Urry \& Padovani 1995).  Blazars are  dominated by rapid and large amplitude variability (e.g., Raiteri et al. 2012; Sobolewska et al. 2014). Multi-wavelength observations show that their broad spectral energy distributions (SED) from the radio to the $\rm \gamma$-rays bands generally exhibits two humps, indicating two components. It is generally accepted that the low-energy component that extends from radio up to ultraviolet, or in some extreme cases to a few keV X-rays (Costamante et al. 2001), is produced by synchrotron radiation from relativistic electrons in the jet (Urry 1998), though the origin of the high-energy component that covers the X-ray and $\rm \gamma$-ray energy regime remains an open issue. There are two kinds of theoretical models describing the high-energy photon emission in these blazars, the  leptonic and the hadronic model. In the leptonic model scenarios, the high-energy component is probably  produced from inverse Compton (IC) scattering of the relativistic electrons either on the synchrotron photons (e.g., Maraschi et al. 1992; Bloom \& Marscher 1996; Mastichiadis \& Kirk 1997; Konopelko et al. 2003) and/or on some other photon populations (e.g., Dermer et al. 1992; Dermer \& Schlickeiser. 1993; Sikora et al. 1994; Ghisellini \& Madau 1996; B$\rm \ddot{o}$ttcher \& Dermer 1998). In contrast, the hadronic model argues that high-energy $\gamma$ rays are produced by either proton synchrotron radiation in high enough magnetic fields (Aharonian 2000; M$\ddot{\rm u}$cke \& Protheroe 2001; M$\ddot{\rm u}$cke et al. 2003; Petropoulou 2014), or mesons and leptons through the cascade initiated by proton-proton or proton-photon interactions (e.g., Mannheim \& Biermann 1992; Mannheim 1993; Pohl \& Schlickeiser 2000; Atoyan \& Dermer  2001).

The imaging atmospheric Cherenkov Telescopes (IACTs) have so far detected  about 50 very high-energy (VHE; $E_{\gamma}>$100 GeV) $\gamma$-ray blazars with redshifts up to $z\sim 0.6$\footnote{http://tevcat.uchicago.edu}. It is believed that the primary TeV photons propagating through intergalactic space should be attenuated due to their interactions with the extragalactic background light (EBL) to produce electron-positron ($e^{\pm}$) pairs (e.g., Nikishov 1962; Gould \& Schreder 1966; Stecker et al. 1992; Ackermann et al. 2012; Abramowski et al. 2013; Dwek \& Krennrich 2013; Sanchez et al. 2013). However, the observed spectra from distant blazars do not show a sharp cutoff at energies around 1 TeV, which would be expected from  simple $\gamma$-ray emission models with a correction for EBL absorbtion (e.g., Stecker et al. 2006; Aharonian et al. 2006a; Costamante et al. 2008; Acciari et a. 2009; Abramowski et al. 2012). Excluding a large uncertainty  in the measured redshifts and in the spectral indices (Costamante 2013) and excluding the lower levels of EBL (Aharonian et al. 2006b; Mazin \& Raue 2007; Finke \& Razzaque 2009), the observed spectral hardening assumes either that there are axion-like particles (de Angelis et al. 2007; Simet et al. 2008; Sanchez-Conde et al. 2009) or a Lorentz invariance violation (Kifune 1999; Protheroe \& Meyer 2000). Alternatively, now that the AGN jets are believed to be one of the most powerful sources of cosmic rays, as long as the intergalactic magnetic fields (IGMF) deep in the voids are less than a femtogauss, the point images of distant blazars, which are produced by the interaction of the energy protons with the background photons along the line of sight, should be observed by IACTs (Essey et al. 2011a). In this scenario, the hard TeV spectra can be produced by the cascade emission from high-energy protons propagating through intergalactic space (Essey \& Kusenko 2010; Essey et al. 2010; 2011b; Razzaque et al. 2012; Aharonian et al. 2013; Takami et al. 2013; Zheng \& Kang 2013).

In this paper, we study  the possible TeV emission in distant TeV blazars. We argue that the TeV emission in distant TeV blazars is dominated by two components, the internal component with the photon energy around 1TeV  produced by IC scattering of the relativistic electrons on the synchrotron photons (SSC) with a correction for EBL absorbtion and  the external component with the photon energy of more than 1TeV  produced by the cascade emission from high-energy protons propagating through intergalactic space. Generally, the external photons are generated in two types photohadronic interactions process along the line of sight. In the first,  the proton interaction with cosmic microwave background (CMB) photons would produce $e^{\pm}$ pairs, and the pairs would give rise to  electromagnetic cascades. This process is called the Bethe-Heitler pair production ($pe$) process. In the second process,  the proton interaction with EBL photons would produce pions, and the pion decay accompanying the photons. This process is called the photopion production ($p\pi$) process. Although the $pe$ process contribution to the production of secondary photons is illustrated at
the source (Dimitrakoudis et al. 2012; Murase 2012; Murase et al. 2012; Petropoulou 2014; Petropoulou \& Mastichiadis 2015), the high-energy astrophysical interest focuses on the $p\pi$ process along the line of sight because the $pe$ process is not associated with any neutrinos and neutrons (e.g., Inoue et al. 2013; Kalashev et al. 2013). The aim of the present work is to study in more detail the contribution of pairs injected by the $pe$ process along the line of sight to the TeV spectra in distant blazars.

Throughout the paper, we assume the Hubble constant $H_{0}=75$ km s$^{-1}$ Mpc$^{-1}$, the matter energy density $\Omega_{\rm M}=0.27$, the radiation energy density$\Omega_{\rm r}=0$, and the dimensionless cosmological constant $\Omega_{\Lambda}=0.73$.

\section{The model}
\label{sec:model}
We calculate the spectra of the internal photon component within the traditional SSC model frame, and we focus on an overlooked Bethe-Heitler pairs cascade process along the line of sight for the spectra of the external photon component. In the following, we give a brief description of the model for possible TeV emission in a distant TeV blazar.
\subsection{Internal photon component}
We assume that a single homogeneous spherical radiation region filled with extreme-relativistic electrons, in which there is a randomly originated homogeneous magnetic field and constant electron number density. We adopt a broken power-law function with a sharp cut-off to describe the electron energy distribution in the radiation region,
\begin{equation}
N_{e, in}(E_{e, in})=
\left\{
\begin{array}{ll}
K_{0}E_{e, in}^{-n_{1}}\;,\mbox{$E_{e, in, min}\le E_{e, in}\le E_{e, in, b}$}\;;\\
K_{1}E_{e, in}^{-n_{2}}\;,\mbox{$E_{e, in, b}\le E_{e,in}\le E_{e,i n, cut}$}\;,
\end{array}
\right.
\label{eq:1}
\end{equation}
where $E_{e, in}$ is the energy of electron in the internal emission region and $K_{1}=K_{0}E_{e, in, b}^{(n_{2}-n_{1})}$. Based on the above electron number density $N_{e, in}(E_{e, in})$, we can calculate the synchrotron intensity $I_{s}(E_{\gamma})$ and the intensity of self-Compton radiation $I_{c}(E_{\gamma})$, and then calculate the intrinsic photon spectrum $dN_{\gamma}^{int}(E_{\gamma})/dE_{\gamma}$ at the observer's frame (e.g., Katarzynski et al. 2001; Zheng \& Zhang 2011; Zheng \& Kang 2013). Taking into account the absorption effect, the flux density observed at the Earth becomes
\begin{equation}
\frac{dN_{\gamma}^{in}(E_{\gamma})}{dE_{\gamma}}=\frac{dN_{\gamma}^{int}(E_{\gamma})}{dE_{\gamma}}\exp[-\tau(E_{\gamma},z)]\;,
\end{equation}
where $\tau(E_{\gamma},z)$ is the absorption optical depth due to interactions with the EBL (e.g., Kneiske et al. 2004; Dwek \& Krennrich 2005; Franceschini et al. 2008). In our calculation, we use the absorption optical depth which is deduced by the EBL model in Dwek \& Krennrich (2005).
\subsection{External photon component}
Active galactic nuclei are expected to accelerate cosmic rays to energies up to $\sim 10^{11}$ GeV during rare bursts or flares (Dermer et al. 2009; Murase \& Takami 2009). The high-energy cosmic rays with energies below the Greisen-Zatsepin-Kuzmin (GZK) cutoff of about $5\times 10^{10}$ GeV can propagate cosmological distances without significant energy loss (Greisen 1966; Zatsepin \& Kuzmin 1966). However, with a small probability, these protons should interact with the CMB photons and trigger electromagnetic cascades.  The present work differs from the earlier studies on which the electromagnetic cascades were modeled using a standard Monte Carlo approach (e.g., Essey et al. 2010). We employ a Bethe-Heitler pair-creation rate in the $\delta$-functional approximation and we concentrate on the protons with energy below the GZK cutoff, which would propagate through cosmological distances.
\subsubsection{Bethe-Heitler pairs production}
We consider an isotropic target photon field with energy $E_{s}=\epsilon_{s} m_{e}c^{2}$ interaction on the proton with energy $E_{p}=\gamma_{p}m_{p}c^{2}$. We let $N_{\rm p}(E_{\rm p})$ and $N_{\rm ph}(E_{\rm s})$ be functions characterizing the energy distributions of protons and soft photons and the collision rate in the unit of $s^{-1}$ can be given by Romero \& Vila (2008)
\begin{equation}
 w_{p\gamma,e}(E_{p})=\frac{m_{p}^{2}m_{e}^{2}c^{9}}{2E_{p}^{2}}\int^{\infty}_{E_{s,min}}\frac{N_{ph}(E_{s})}{E_{s}^{2}}dE_{s}
\int^{\epsilon_{s, max}^{'}}_{\epsilon_{s, min}^{'}}\sigma_{p\gamma,e}(\epsilon^{'}_{s})\epsilon^{'}_{s} d\epsilon^{'}_{s}\;,
\end{equation}
where $\epsilon^{'}_{s}=\gamma_{p}\epsilon_{s}(1-\beta_{p}\cos\theta)$ is the energy of the photon in the rest frame of the proton with the angle between the proton and photon directions $\theta$, and the proton's velocity $\beta_{p}$ in units of $c$. Because  the Bethe-Heitler pairs production are with a threshold energy of $\sim$1 MeV for the photon in the proton rest frame, we can obtain $\epsilon_{s, min}^{'}=1 \rm MeV/$$(m_{e}c^{2} )\sim2$, $E_{s,min}=1/2\gamma_{p}$ MeV (e.g., Mastichiadis \& Kirk 1995). A head-on collision gives the maximum energy of the photon in the rest frame of the proton $\epsilon_{s, max}^{'}=2\gamma_{p}\epsilon_{s}$. The energy distribution of soft photons is described by \begin{equation}
N_{ph}(E_{s})=\frac{1}{\pi^{2}(\hbar c)^{3}}\frac{E_{s}^{2}}{\exp(E_{s}/kT)-1}\;,
\end{equation}
where the typical temperature $T=(1+z)2.73$ K with redshift $z$ and $k$ is the Boltzmann constant. The total cross section $\sigma_{p\gamma,e}(\epsilon^{'}_{s})$ for which we use the Racah formula as parameterized by Maximon (1968). For $2\le\epsilon^{'}_{s}\le4$, the expansion is given with a fractional error less than $1.1\times10^{-3}$ (Maximon 1968),
\begin{eqnarray}
\sigma_{p\gamma,e}(\epsilon^{'}_{s})&\simeq&\frac{2\pi}{3}\alpha_{f}r_{0}^{2}Z^{2}\left(\frac{\epsilon^{'}_{s}-2}{\epsilon^{'}_{s}}\right)^{2}\nonumber\\ &\times&\left(1+\frac{1}{2}\eta+\frac{23}{40}\eta^{2}+\frac{37}{120}\eta^{3}+\frac{61}{192}\eta^{4}\right)\;,
\end{eqnarray}
where $\eta=(\epsilon^{'}_{s}-2)/(\epsilon^{'}_{s}+2)$, $\alpha_{f}$ is the fine structure constant, $r_{0}$ is the classical electron radius, and $Z$ is the Atomic number of the target atom. In the larger photon energy regime, $\epsilon^{'}_{s}\ge 4$, the expansion can be written with a fractional error less than $4.4\times10^{-5}$ as (e.g., Chodorowski et al. 1992)
\begin{eqnarray}
&\sigma&_{p\gamma,e}(\epsilon^{'}_{s})\simeq\alpha_{f}r_{0}^{2}Z^{2}\biggl\{\frac{28}{9}\ln 2\epsilon^{'}_{s}-\frac{218}{27} +\left(\frac{2}{\epsilon^{'}_{s}}\right)^{2}\nonumber\\&\times&\biggl[6\ln 2\epsilon^{'}_{s}-\frac{7}{2}+\frac{2}{3}\ln^{3}2\epsilon^{'}_{s}-\ln^{2}2\epsilon^{'}_{s}-\frac{1}{3}\pi^{2}\ln 2\epsilon^{'}_{s}\nonumber\\&+&2\zeta(3)+\frac{\pi^{2}}{6}\biggr]-\left(\frac{2}{\epsilon^{'}_{s}}\right)^{4}\left(\frac{3}{16}\ln 2\epsilon^{'}_{s}+\frac{1}{8}\right)\nonumber\\&-&\left(\frac{2}{\epsilon^{'}_{s}}\right)^{6}\left(\frac{9}{9\cdot256}\ln 2\epsilon^{'}_{s}-\frac{77}{27\cdot512}\right)\biggr\}\,,
\end{eqnarray}
where $\zeta(3)\simeq1.20206$. Now the Monte Carlo simulation shows that the mean inelasticity can be approximated by its values at the threshold $K_{p\gamma,e}(\epsilon^{'}_{s})=2m_{e}/m_{p}$ (Mastichiadis et al. 2005). Therefore, the Bethe-Heitler pair-creation rate in the $\delta$-functional approximation is then given (Romero \& Vila 2008) as
\begin{eqnarray}
Q_{e}(E_{p\gamma,e})&=&2\int_{E_{p,min}}^{E_{p,max}}N_{p}(E_{p})w_{p\gamma,e}(E_{p})\delta\left(E_{p\gamma,e}-\frac{m_{e}}{m_{p}}E_{p}\right)dE_{p}\nonumber\\
&=&2\frac{m_{p}}{m_{e}}N_{p}\left(\frac{m_{p}}{m_{e}}E_{p\gamma,e}\right)w_{p\gamma,e}\left(\frac{m_{p}}{m_{e}}E_{p\gamma,e}\right)\;.
\end{eqnarray}
We parameterize the proton injection spectrum by a constant power-law exponent $\alpha$ and maximum energy $E_{p,max}$:
\begin{equation}
N_{p}(E_{p})=N_{0}E_{p}^{-\alpha}\exp\left(-\frac{E_{\rm p}}{E_{\rm p, max}}\right)\;, \mbox{$E_{p,min}\le E_{p}\le E_{p,max}$}\;.
\end{equation}
Where the normalization coefficient $N_{0}$ is determined from the jet power $L_{p}$ of the protons. The observed data on the most distant sources shows that an effective luminosity of a single AGN in cosmic rays above $10^{7}$ GeV was in the range $L_{eff}\sim10^{47}-10^{49}$ erg s$^{-1}$. Since the VLBI observation shows that the value of the Doppler factor $\delta$ is less than 10 (Wu et al. 2007), the flux can be beamed with the beaming factor $f_{beam}\sim10-10^{3}$, which increases the flux of protons from a blazar with a jet pointing in the direction of Earth (Essey \& Kusenko, 2010)
\begin{equation}
L_{eff}\sim10^{2}f_{p}\times(\frac{f_{beam}}{100})L_{p}\;,
\end{equation}
where $f_{p}\leq1$ is the fraction of protons.

\subsubsection{Inverse Compton scattering from Bethe-Heitler pairs}
Assuming that the energy of the Bethe-Heitler pairs is lost  through interaction with the CMB photons at the cosmological distance D, we can deduce energy distribution of the secondary electrons or positrons from the well-known energy distribution equation
\begin{equation}
\frac{dN_{e, ex}(E_{e, ex})}{dE_{e, ex}}=\left(\frac{dE_{e, ex}}{dt}\right)^{-1}\int_{E_{e, ex}}^{E_{p\gamma,e,max}}Q_{e}(E_{p\gamma,e})dE_{p\gamma,e}\;,
\end{equation}
where $dE_{e, ex}/dt=4\sigma_{T}E_{e, ex}^{2}/(3m_{e}^{2}c^{3}u_{CMB})$ is the Bethe-Heitler pairs energy loss rates with Thomson cross section $\sigma_{T}$ and CMB energy density $u_{CMB}$. Then, the photon spectrum of the inverse Compton scattering is given by
\begin{equation}
\frac{dN_{\gamma}^{ex}(E_{\gamma})}{dE_{\gamma}}=\int_{E_{e, ex, min}}^{\infty}\frac{dN_{e, ex}(E_{e, ex})}{dE_{e, ex}}\biggl[\frac{d^{2}N_{ph}(E_{s})}{dE_{s}dt}\biggr]_{ICS}dE_{e, ex}\;,
\end{equation}
with the spectrum of the inverse Compton scattered photons per electron (Blumenthal \& Gould 1970)
\begin{eqnarray}
\biggl[\frac{d^{2}N_{ph}(E_{s})}{dE_{s}dt}\biggr]_{ICS}&=&\int_{E_{s,min}}^{\infty}\frac{3\sigma_{T}m_{e}^{2}c^{5}}{4E_{e, ex}^{2}}\frac{N_{ph}(E_{s})}{E_{s}}dE_{s}
\times\biggl[2q\ln q\nonumber\\&+&(1+2q)(1-q)+\frac{(\Gamma q)^{2}(1-q)}{2(1+\Gamma q)}\biggr]\;,
\end{eqnarray}
where $q=E_{1}/\Gamma(1-E_{1})$, $\Gamma=4\epsilon_{s}E_{e, ex}/m_{e}c^{2}$, and $E_{1}=E_{\gamma}/E_{e, ex}$.

Therefore, the TeV photon spectrum that is observed at the Earth is given by
\begin{equation}
\frac{dN_{\gamma}(E_{\gamma})}{dE_{\gamma}}=\frac{dN_{\gamma}^{in}(E_{\gamma})}{dE_{\gamma}}+\frac{dN_{\gamma}^{ex}(E_{\gamma})}{dE_{\gamma}}\;.
\end{equation}
\section{Constraint on the energy of protons}
The above scenario depends on the energy of protons. Since the IGMF would cause proton or electron deflection in the voids, and the energetic $\gamma$ rays in the EBL photon fields should be absorbed, we can determine the ranges of a proton's energy through these physical process.

As long as the IGMF $B_{IG}$ is less than 1 femtogauss, the Cosmic ray protons with energies $E_{p}\le E_{GZK}=5\times10^{10}$ GeV can provide an effective transport of the energy over a cosmological distance toward the observer (Essey et al. 2011a). Then they interact with the CMB photons and initiate electromagnetic cascades. On condition that the deflection broadening of the proton beam $\theta_{p}$ in IGMFs is less than the point spread function $\overline{\theta(E)}$ of the detector, the resulting secondary photons are observed as arriving from a point source. The deflection angles dependence on the IGMF can be written (Aharonian et al. 2010)
\begin{equation}
\theta_{p}\simeq0.05~arcmin\left(\frac{10^{9}~GeV}{E_{p}}\right)\left(\frac{B_{IG}}{10^{-15}~G}\right)\left(\frac{l_{c}}{Mpc}\frac{d}{Gpc}\right)^{1/2}\;,
\end{equation}
where $l_{c}$ is a correlation length of the random fields and $d\sim cz/H_{0}$ is luminosity distance. We consider a VHE blazar with redshift $z$, and the typical point spread function of the HESS telescope array, $\overline{\theta(E)}\simeq3.0$ arcmin. The constraint condition $\theta_{p}\leq\overline{\theta(E)}$ could result in $E_{p}\geq E_{p,low}=3.3\times10^{7}z^{1/2}(l_{c}/1Mpc)^{1/2}(B_{IG}/10^{-15}~G)$ GeV.

\begin{figure}
\vspace{0.0cm}
\label{Fig:1}
\centering
\includegraphics[angle=0,width=9cm]{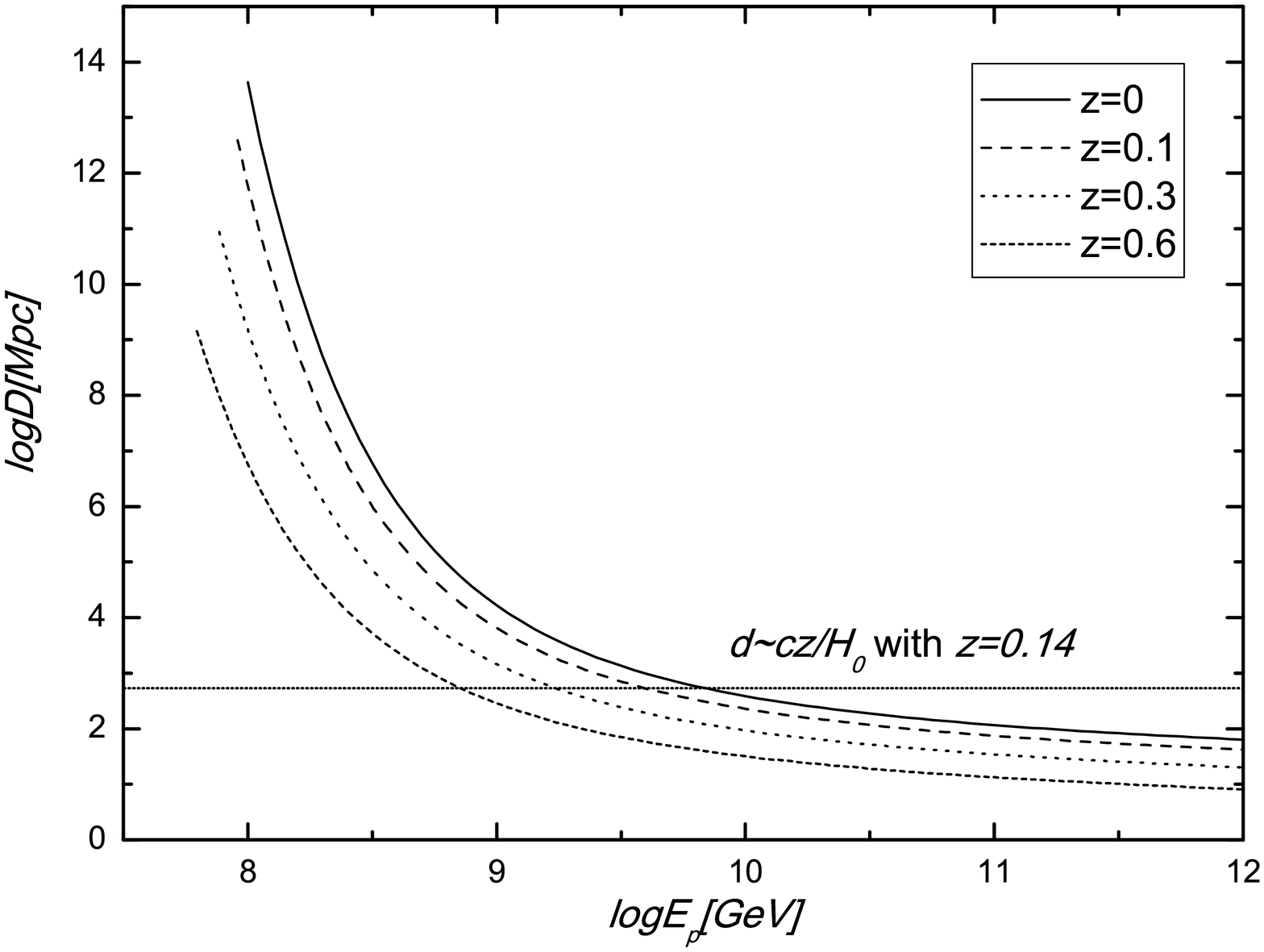}
\caption{Secondary emission region as a function of energy and redshift. The distance to the source that is estimated by using Hubble's law with z=0.14 is also exhibited as a short dotted line.}
  \label{Fig:1}
\end{figure}

Alternatively, VHE $\gamma$-ray photon emission from the region is attenuated by photons from the EBL (e.g., Kneiske et al. 2004; Dwek \& Krennrich 2005). It is well-known that the $\gamma$-ray zone can be defined by the $\gamma$-ray absorption mean free path $\lambda_{\gamma\gamma}\sim190(n_{IR}/0.01cm^{-3})^{-1}$ Mpc with the infrared photon number density $n_{IR}$ (e.g., Venters 2010). As long as the magnetic field is significantly small, an extremely efficient cascade development could enlarge the $\gamma$-ray zone, typically $\lambda_{\gamma,eff}\sim2-3\lambda_{\gamma\gamma}$ (Aharonian et al. 2013). In this scenario, we argue that if the  Bethe-Heitler cascades contribution to the observed hard spectra behavior of blazars is true, the secondary $\gamma$-ray emission region $D$ should satisfy a relation $d-D\leq\lambda_{\gamma,eff}$. In our work, due to the cascades the electrons are the main origin of the $pe$ process, the adoption $D\simeq\lambda_{p\gamma,e}=c/t_{p\gamma,e}^{-1}(E_{p})$ is justified. Here,
\begin{eqnarray}
t_{p\gamma,e}^{-1}(E_{p})&=&\frac{m_{p}^{2}m_{e}^{2}c^{9}}{2E_{p}^{2}}\int^{\infty}_{E_{s,min}}\frac{N_{ph}(E_{s})}{E_{s}^{2}}d E_{s}\nonumber\\
&\times&\int^{\epsilon_{s, max}^{'}}_{\epsilon_{s, min}^{'}}K_{p\gamma,e}(\epsilon^{'}_{s})\sigma_{p\gamma,e}(\epsilon^{'}_{s})\epsilon^{'}_{s} d\epsilon^{'}_{s}\;
\end{eqnarray}
is the cooling rate of the protons in CMB photon fields (Romero \& Vila 2008). Obviously, in the case of a broad energy distribution of protons, the main contribution to the secondary $\gamma$-ray flux comes from some energetic proton ranges in which the secondary $\gamma$-ray emission region is comparable to the distance of the source, i.e., $D\sim d$. We show the secondary emission region as a function of energy and redshift in Fig.\ref{Fig:1}. It can be seen that the characteristic energy of proton $E^{*}_{p}\sim6.3\times10^{9}$ GeV should be found in Fig.\ref{Fig:1} as the point where the distance to the source is equal to the mean free path of protons at the present epoch (z=0). The above issues give a strong constraint on the range of energy of the  protons from $E_{p,min}=E_{p,low}$ to $E_{p,max}=min[E^{*}_{p},E_{GZK}]$.

\section{Apply to 1ES 0229+200}
\label{sec:apply}
1ES 0229+200 resides in an elliptical host galaxy at a redshift of $z=0.1396$ (Woo et al. 2005). The source has been classified as a high-frequency peaked BL BLac object (HBL) due to its X-ray to radio flux ratio (Giommi et al. 1995). As a special source, the spectrum of 1ES 0229+200 was measured extended up to 10 TeV with a hard archival spectral index at VHE of $2.5\pm0.19$ (Aharonian et al. 2007; Aliu et al. 2014). Compiling with the VERITAS measured spectrum averaged over all three seasons (Aliu et al. 2014), and four years
of Fermi-LAT  observed results\footnote{http://fermi.gsfc.nasa.gov/ssc/data/access/lat/4yr$\_$catalog/}, we reanalyze the  spectrum of 1ES 0229+200 assuming a relation $dN_{\gamma}(E_{\gamma})/dE_{\gamma}\propto E_{\gamma}^{-\Gamma}$ in Fig. \ref{Fig:2}. The results of linear regression analysis are listed in table 1. It can be seen that the spectrum should be fitted by a power-law relation with the spectra index $\Gamma_{MeV-GeV, obs.}=1.93\pm0.06$ in the MeV-GeV energy band, and $\Gamma_{TeV, obs.}=2.27\pm0.13$ in the TeV energy band. Now that the model argues that the TeV emission in distant TeV blazars is dominated by two components, the external TeV photons could significantly repair the EBL attenuation, and leave a hard spectra to 100 TeV energy band. Using the model in $\S$2, we should expect to reproduce the hard TeV $\gamma$-rays spectra.

\begin{table}
\caption{Linear regression analysis$^{a}$ of the observed spectrum of 1ES 0229+200 in the MeV-TeV energy band}
\centering
\begin{tabular}{ccc}
\hline\hline
photon energy & $ MeV-GeV $ & $\gtrsim$1TeV\\
\hline\\
$\Gamma(\sigma)$  & $1.93\pm0.06$   & $2.27\pm0.13$ \\
   r               & 0.99            & 0.98 \\
   N               & 13              & 13 \\
   p               & $<$0.0001         & $<$0.0001 \\
\hline\\
\end{tabular}
\label{Table:1}
\flushleft{NOTE: $^{a}$ The linear regression is obtained by considering the photon energy $E_{\gamma}$ to be the independent variable and assuming a relation $ dN_{\gamma}(E_{\gamma})/dE_{\gamma}\propto E_{\gamma}^{-\Gamma}$; N is the number of points, r is the correlation coefficient, and p is the chance probability.}
\end{table}

\begin{figure}
\vspace{0.0cm}
\label{Fig:2}
\centering
\includegraphics[angle=0,width=9cm]{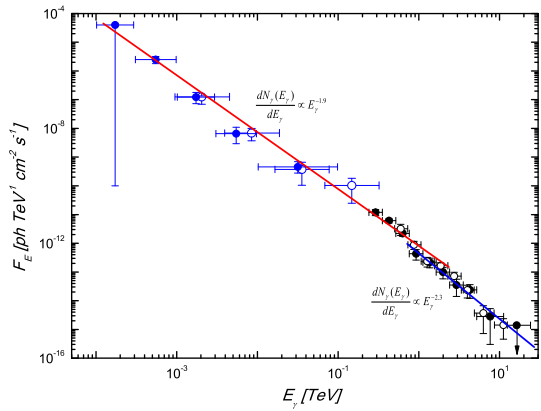}
\caption{Observed photon spectrum of 1ES 0229+200.  The solid lines are the linear regression results in the MeV-GeV and  TeV energy bands. The Fermi-LAT observed data come from the second catalog (blue open circle) and third catalog (blue solid circle). The HESS observed data (black open circle) come from Aharonian et al. (2007) and VERITAS observed data (black solid circle) come from Aliu et al. (2014).}
  \label{Fig:2}
\end{figure}

In order to do so, first we search for the internal $\gamma$-ray component in the one-zone homogeneous SSC frame. Assuming the density normalization $K_{0}=5.57\times10^{3}~erg^{-1}~cm^{-3}$, we calculate the high-energy electron distribution with a broken power law between $E_{e, in, min}=4.34$ GeV and $E_{e, in, cut}=4.5\times10^{4}$ GeV with a break at $E_{e, in, b}=4.49\times10^{2}$ GeV, where the energy index of the particles between $E_{e, in, min}$ and $E_{e, in, b}$ is set to $n_{1}=2.2$ and the energy index of the particles between $E_{e, in, b}$ and $E_{e, in, cut}$ is set to $n_{2}=4.0$, The parameters are used as follows: the magnetic field strength is $B=0.06$ G, the emission
region size is $R_{e}=2.77\times10^{16}$ cm, and the Doppler factor is $\delta=10.7$. We assume that relativistic electrons are in the steady state during the observational epoch. Therefore, we can
calculate the TeV $\gamma$-ray spectrum in the one-zone homogeneous SSC frame using the broken power-law electron spectrum.

We now consider the external TeV photons production along the line of sight. We argue that the sites of electron acceleration may also accelerate protons  in the jet. If the acceleration site is full of electrically neutral nonthermal plasma, the proton-to-electron ratio depends on the spectral indices of injection electron and proton spectra (e.g., Schlickeiser 2002; Persic \& Rephaeli 2014). As a simple case, we do not specify the proton-to-electron ratio and leave it as a free parameter with $L_{p}$ and $L_{e}$. Assuming the protons are accelerated to extra-relativistic energy in the acceleration region through some mechanisms and then escape from the region without energy loss. We calculate the Bethe-Heitler pairs distribution along the line of sight with a proton injection spectrum between $E_{\rm p, min}$ and $E_{\rm p, max}$, where, in this case, the correlation length of the random fields is set to $l_{c}=1$ Mpc, the IGMF is set to $B_{IG}=10^{-15}$ G (e.g., Tavecchio et al. 2010), and the power-law exponent $\alpha=2.0$ is adopted by fitting to the ultra high-energy cosmic rays data at the lower energies (Berezinsky et al. 2006). Since the typical luminosity range of AGNs are in $10^{45}-10^{47} ~erg~s^{-1}$ (e.g., Ghisellini et al. 2014), in our calculation we set the jet power of protons $L_{p}=0.3\times10^{46}~erg~s^{-1}$. We assume that the energy of the Bethe-Heitler pairs is lost  through interaction with the CMB photons at the cosmological distance D during the observational epoch. Therefore, we can calculate the external TeV $\gamma$-ray spectrum along the line of sight using the injection electron spectrum.

We show the predicted internal component spectrum, external component spectrum, and the superposed model spectral in Fig.\ref{Fig:3}. For comparison, the differential sensitivity curve for the 50-hour observation with H.E.S.S. I\footnote{http://www.mpi-hd.mpg.de/hfm/HESS/pages/home/proposals/}, the 50-hour differential sensitivity goal of the Cherenkov Telescope Array (CTA, Actis et al. 2011), the differential sensitivity curve for the one-year observation with the Large High Altitude Air Shower Observatory (LHAASO, Cui et al. 2014), and the multi-wavelength observed data of 1ES 0229+200 (Aliu et al. 2014) are also shown. All of the physical parameters of the internal component and external component spectra are listed in table 2. It can be seen that  1) the observed spectrum properties of 1ES 0229+200, especially the TeV $\gamma$-ray tail of the observed spectra, were  reproduced in our model; 2) an expected TeV $\gamma$-ray spectrum with photon energy $>$1 TeV of 1ES 0229+200 should be comparable with the 50-hour sensitivity goal of the CTA and the differential sensitivity curve for the one-year observation with the LHAASO.

In order to intensively examine the hard spectra properties of 1ES 0229+200 in the TeV energy bands. We show the model spectra with different EBL absorption optical depths in Fig. \ref{Fig:4}. Now that the predicted spectral shape of external photons is not very sensitive to the variations in the proton spectrum index $\alpha$ (e.g., Essey et al. 2010), we do not take into account the effect of $\alpha$ on the model spectra. It is indicated that, dependent of the EBL absorption optical depth, the photon spectra with the photon energy between 0.1 TeV and 10 TeV could be described by a relation $ dN_{\gamma}(E_{\gamma})/dE_{\gamma}\propto E_{\gamma}^{-\Gamma}$ with a small photon spectral index $1.5<\Gamma<2.5$, and even  flatter in a higher energy band. As a special case with the LLL EBL model in Dwek \& Krennrich (2005), in Fig.\ref{Fig:5} we compare the predicted differential photon spectrum in the $\gtrsim$ 1TeV $\gamma$-ray energy bands (blue solid curve) with observed data of 1ES 0229+200 at HESS (Aharonian et al. 2007) and VERITAS (Aliu et al. 2014) energy bands. Our result shows that in this special case, superposed model spectra with the photon energy between 1 TeV and 10 TeV could be described by a photon spectral index $\Gamma=2.4$. This is in agreement with the result that is found by reanalysis of the  spectrum of 1ES 0229+200.

\begin{table}
\caption{Physical parameters of the model spectra}
\centering
\begin{tabular}{llcll}
\hline\hline
\multicolumn{2}{c}{Internal Component} & & \multicolumn{2}{c}{External Component}\\
\cline{1-2} \cline{4-5} \\
 Parameters  & value  & & parameters  &  value \\
\hline\\
$K_{0}~[\rm erg^{-1}~cm^{-1}]$  &  $5.57\times10^{3}$  &    &  $E_{\rm p,max}~[\rm GeV]$  & $6.3\times10^{9}$ \\
$n_{1}$                         &    2.2    &    &  $\alpha$  & 2.0   \\
$n_{2}$                         &    4.0    &    &  $\rm B_{IG}~[\rm G]$       & $1.0\times10^{-15}$ \\
$E_{e, in, min}~[\rm GeV]$      &  4.34     &    &   $l_{c}~[\rm Mpc]$          & 1.0           \\
$E_{e, in, b}~[\rm GeV]$        &   449      &    &  $L_{p}~[\rm erg~s^{-1}]$   & $0.3\times10^{46}$         \\
$E_{e, in, cut}~[\rm GeV]$      &  $4.5\times10^{4}$   &    & D~[Gpc] & 1.0  \\
$B~[\rm G]$                     &   0.06           &    & --&--  \\
$\delta$                        &   10.7           &    &  --  & -- \\
$R_{e}~[\rm cm]$                &  $2.77\times10^{16}$      &    &  --   & -- \\
\hline\\
\end{tabular}
\label{Table:2}
\end{table}

\begin{figure}
\vspace{0.0cm}
\label{Fig:3}
\centering
\includegraphics[angle=0,width=9cm]{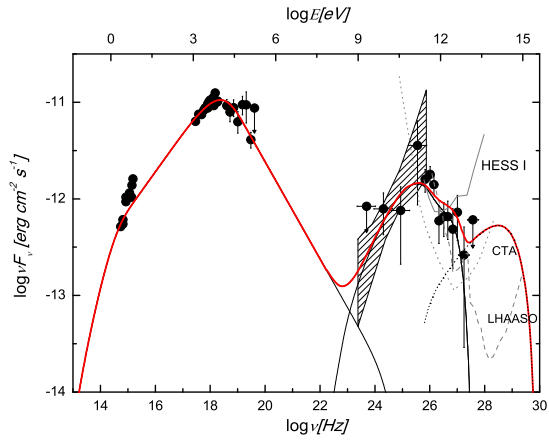}
\caption{Multiwavelength SED of the distant TeV blazar 1ES 0229+200. The solid black curve represents the internal component spectra with the absorption optical depth that is deduced by the LLL EBL model in Dwek \& Krennrich (2005), and the dotted black curve represents the external component spectra. The superposed model spectrum is plotted as a red curve. The observed data come from Aliu et al. (2014) (solid circle). The shaded area shows the Fermi upper bounds at the   99\%\ confidence level. The differential sensitivity curve for the 50-hour observation with H.E.S.S. I (http://www.mpi-hd.mpg.de/hfm/HESS/pages/home/proposals/, gray solid curve), the 50-hour differential sensitivity goal of the CTA (Actis et al. 2011, gray dashed curve), and the differential sensitivity curve for the one-year observation with LHAASO (Cui et al. 2014, gray dotted curve) are also shown.}
  \label{Fig:3}
\end{figure}

\begin{figure}
\vspace{0.0cm}
\label{Fig:4}
\centering
\includegraphics[angle=0,width=9cm]{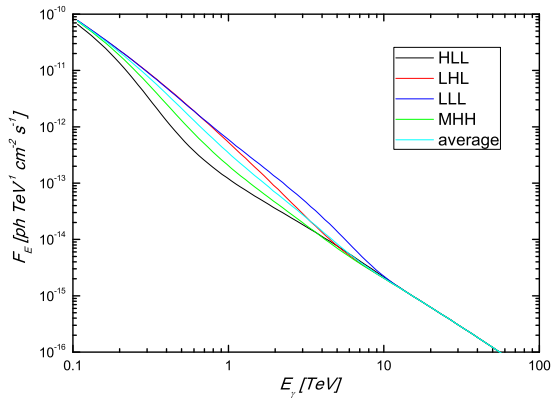}
\caption{Model spectra with different EBL absorption optical depths. It is indicated that, dependent of the EBL absorption optical depth, the photon spectra with the photon energy between 0.1 TeV and 10 TeV could be described by a relation $ dN_{\gamma}(E_{\gamma})/dE_{\gamma}\propto E_{\gamma}^{-\Gamma}$ with a small photon spectral index $1.5<\Gamma<2.5$, and even  flatter in a higher energy band.}
  \label{Fig:4}
\end{figure}

\begin{figure}
\vspace{0.0cm}
\label{Fig:5}
\centering
\includegraphics[angle=0,width=9cm]{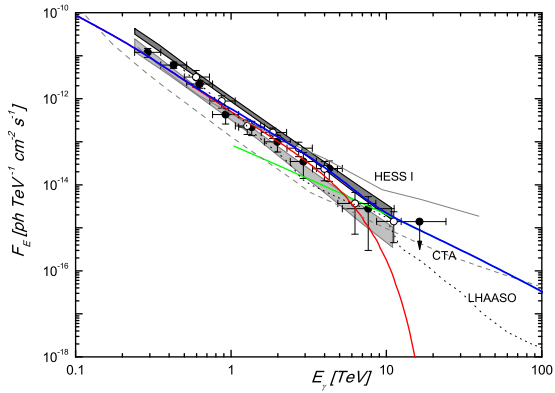}
\caption{Comparisons of predicted TeV photon spectra with observed data of the distant TeV blazar 1ES 0229+200. The upper and lower shaded regions show the spectral shape during the flaring and low periods. The red curve represents the internal component spectra with the absorption optical depth that is deduced by the LLL EBL model in Dwek \& Krennrich (2005), and the green curve represents external component spectra. The superposed model spectrum is plotted as a blue curve. The HESS Observed data (open circle) come from Aharonian et al. (2007) and VERITAS observed data (solid circle) come from Aliu et al. (2014). The integral sensitivity curve for the 50-hour observation with H.E.S.S I (http://www.mpi-hd.mpg.de/hfm/HESS/pages/home/proposals/, gray solid curve), the 50-hour integral sensitivity goal of the CTA (Actis et al. 2011, gray dashed curve), and the integral sensitivity curve for the one-year observation with LHAASO (Cui et al. 2014, gray dotted curve) are also shown.}
  \label{Fig:5}
\end{figure}

\section{Discussion and conclusions}
\label{sec:discussion}
As an open issue, very high-energy $\gamma$-ray measurements of distant TeV blazars can be  explained by  TeV spectra induced by ultra high-energy cosmic rays (Essey \& Kusenko 2010; Essey et al. 2010; 2011b; Murase 2012; Murase et al. 2012; Razzaque et al. 2012; Takami et al. 2013; Zheng, et al. 2013). In this paper, We develop a model for a possible TeV emission in distant TeV blazars. The aim of the present work is to study in greater detail the contribution of pairs injected by the $pe$ process along the line of sight to the TeV spectra in distant blazars. In the model, the TeV emission in distant TeV blazars is dominated by two mixed  components:  the first is the internal component where the photon energy around 1 TeV is produced by IC scattering of the relativistic electrons on the synchrotron photons (SSC) with a correction for EBL absorbtion, and the second is the external component where the photon energy more than 1 TeV is produced by the cascade emission from high-energy protons propagating through intergalactic space. Assuming a suitable model parameters, we apply the model to observed spectra of distant TeV blazars of 1ES 0229+200. Our results show that 1) the observed spectrum properties of 1ES 0229+200, especially the TeV $\gamma$-ray tail of the observed spectra, could be reproduced in our model and 2) an expected TeV $\gamma$-ray spectrum with photon energy $>$1 TeV of 1ES 0229+200 should be comparable with the 50-hour sensitivity goal of the CTA and the differential sensitivity curve for the one-year observation with LHAASO. We argue that  strong evidence for the Bethe-Heitler cascades along the line of sight as a plausible origin of hard spectra in distant TeV blazars could be obtained from VHE observations with CTA and LHAASO.

The present work differs from the earlier studies that assume that the pairs cascade process induced by ultra high-energy cosmic rays   occurs at the source (e.g., Murase 2012; Murase et al. 2012; Petropoulou \& Mastichiadis 2015). We concentrate on the protons with energy below the GZK cutoff, which would propagate through cosmological distances. We argue that the outflows of the jets from AGNs are likely to contain coherent magnetic fields aligned with the jet, so that the accelerated protons remain in the scope of the initial jet rather than getting deflected. Since the $pe$ process takes place outside the galaxy clusters of both the observer and the source, the cluster magnetic fields are irrelevant to this issue. Although we expect larger fields in the filaments and wall, only the IGMF present deep in the voids along the line of sight is important (Essey \& Kusenko 2010). Within the host galaxy, the propagated directions of the protons could be changed by the galactic magnetic fields, the broadening of the image due to deflections in it should less than $\Delta\theta_{max}\sim r/D_{source}$ (Essey et al. 2010), where $r$ is the size of the host galaxy, and $D_{source}$ is the distance to the host galaxy. Furthermore, the possible thin walls of magnetic fields that might intersect the line of sight could not cause a deflection of more than $\Delta\theta\sim h/D_{wall}$ (Essey et al. 2010), where $h$ is the wall thickness and $D_{wall}$ is the distance to the wall. The model also does not take into account both the $\gamma$-ray photon spectrum and the pairs cascade from the decay of pions $\pi^{0}$, $\pi^{+}$, and $\pi^{-}$. We argue that the characteristic energy of both decay induced photons $E_{\gamma}\sim0.1E_{p}>10^{3}$ TeV and pairs cascade from the decay of pions induced photons $E_{\gamma}\sim(0.05E_{p}/m_{e}c^{2})^{2}E_{CMB}>10^{4}$ TeV with IGMF $B_{IG}=10^{-15}$ G and correlation length of the random fields $l_{c}=1$ Mpc. This energy is far from the TeV energy band.

It is noted that  small photon indices are not easy to achieve in traditional leptonic scenarios, although the stochastic acceleration model (Lefa et al. 2011) and the leptohadronic model (Cerruti et al. 2015) can also  explain the spectral hardening of TeV blazars because radiative cooling tends to produce particle energy distributions that are always steeper than $E^{-2}$. The above distribution results in a TeV photon index $\Gamma_{TeV,int}\geq1.5$, and even  steeper at VHE due to the suppression of the Klein-Nishina effects  of the cross-section (e.g., Chiang \& B$\rm \ddot{o}$ttcher, 2002). Even when the absorption effect by the lowest level EBL is used, the emitted spectra still tend to be steeper with an observed photon index $\Gamma_{TeV,obs}\geq 2.5$ (e.g., Aharonian et al. 2007; Dwek, \& Krennrich 2012). \bf {Because the Bethe-Heitler pair-creation rate is smoothed in the model, we argue that the spectral shape of the external component spectrum is not very sensitive to the proton injection spectrum (Essey et al. 2010); it is determined primarily by the spectrum of the CMB photons and the Bethe-Heitler pairs energy loss process, which result in hard TeV photons spectra.} \LEt{because the sentence is hardly understood, we rephrase the sentence.}

Alternatively, the secondary $e^{\pm}$ pairs that are produced by $\gamma+\gamma\to e^{+}+e^{-}$ pair creation generate a new $\gamma$-ray component through IC scattering of these $e^{\pm}$ pairs against target photons of the CMB, initiating an electromagnetic cascade if the produced $\gamma$ ray is subsequently absorbed (e.g., Dai et al. 2002; Fan et al. 2004; Yang et al. 2008; Neronov et al. 2012). In order to reproduce a hard spectra, we include an external $\gamma$-ray component with photon energy around $\sim10-100$ TeV. Using the photon energy of the external $\gamma$-ray component, we could estimate the boosting energy through the IC process of $E_{\gamma}\sim\gamma_{e^{\pm}}^{2}\epsilon_{CMB}\sim0.01-1.0$ TeV. In this view, the resultant secondary photons should contribute to the TeV $\gamma$-ray flux and we expect to find a complex spectrum around 1 TeV. However, ultra high-energy cosmic rays with $E_{p}\sim10^{19}$ eV have  energy loss paths $\lambda_{p\gamma,e}\sim$1 Gpc for Bethe-Heitler pair production, whereas 10-100 TeV $\gamma$-ray only travel $\lambda_{\gamma,eff}\sim$3-200 Mpc before being absorbed by $\gamma\gamma$ pairs production. Now that we assume the cascade emission region $D\sim\lambda_{p\gamma,e}$, the source at a redshift of $z<1.0$ allows the Bethe-Heitler pair to be injected and cascade to such energy far from the source. Thus, some TeV photons can reach the Earth before being attenuated even when $\tau(E_{\gamma}, z)\gg1$, without any spectral shape transformation (Takami et al. 2013). On the other hand, although we focus on spectral information to the possible TeV emission in distant blazars, variability is also another important clue. Murase et al. (2012) argue that the cascade components would not have short variability timescales, since the shortest time scales are $\sim1.0(E_{\gamma}/10~\rm GeV)^{-2}(B_{IG}/10^{-18}~\rm G)^{2}$ yr in the $\gamma$-ray induced cascade case, and $\sim10(E_{\gamma}/10~\rm GeV)^{-2}(B_{IG}/10^{-18}~\rm G)^{2}$ yr in the ultra high-energy cosmic ray induced cascade case.  The above issue suggests that  strong variability is possible in the $\gamma$-ray induced cascade case, and this means  that the cascaded $\gamma$-ray should be regarded as a mixture of attenuated and cascade components, and then the cascade component could be suppressed. Instead,  in the ultra high-energy cosmic ray induced cascade case, the variability could not be found (Takami et al. 2013). The lack of strong evidence of variability in 1ES 0229+200 above the HESS energy band (Aharonian et al. 2007; Aliu et al. 2014) suggests that the ultra high-energy cosmic ray induced cascade component might dominate on the TeV $\gamma$-ray spectrum. Since, in our issue, the TeV emission in distant TeV blazars is dominated by two components, the external TeV photons could significantly repair the EBL attenuation, and leave a hard spectra to 100 TeV energy band.

It is clear that the jet power of protons plays an important role in determining an emission intensity. In our results, in order to obtain the TeV emission of the 1ES 0229+200, we adopt the jet power of protons $L_{p}=0.3\times10^{46}~erg~s^{-1}$. It is well known that the jets of AGN are powered by the accretion of matter onto a central black hole (e.g., Urry \& Padovani 1995). On the assumption that the radiation escapes isotropically from the black hole, the balancing of the gravitational and radiation force leads to the maximum possible luminosity due to accretion $L_{edd}\sim 1.26\times10^{38}M/M_{\odot}~erg~s^{-1}$ (e.g., Dermer \& Menon 2009). When the total emission of an AGN is not super-Eddington, the Eddington luminosity is the maximum power available for the two jets, $P_{jet} \leq L_{edd}/2$. In this view, the required power in protons could easily be provided by the source with mass $M=1.44\times10^{9}M_{\odot}$ (Wagner 2008).

A potential drawback of the model is that the shape of the model spectra in the 1 TeV-10 TeV energy ranges strongly depends on the level of the EBL. As a check, we constrain the shape of the model of spectra depending on a general EBL model. On the basis of the model results, we argue that the predicted TeV spectra properties of the above-mentioned model should  be testable in the near future since the secondary emission process will expect CTA to detect more than 80 TeV blazars in the above 1 TeV energy band (Inoue et al. 2013). We note that the neutrino populations could be expected in $p\gamma$ interactions, and it should be given a  clearer predictive character with the IceCube observations. Unfortunately, the $pe$ process does not contain any neutrino populations.  We defer this possibility to future work. Although our model focuses on the contribution of pairs injected by $pe$ process along the line of sight to the TeV spectra, the $\gamma$-ray induced cascade is also another important scenario for the possible TeV emission in distant blazars (e.g., Vovk et al. 2012; Takami et al. 2013). We leave this possibility to the observation of CTA (Actis et al. 2011), LHAASO(Cao 2010; Cui et al. 2014), HAWC (Sandoval et al. 2009), and HiSCORE (Hampf et al. 2011).



\section*{Acknowledgments}
We thank the anonymous referee for valuable comments and suggestions.
This work is partially supported by the National Natural Science Foundation of China under grants 11463007, U1231203, Science and Technology in support of Yunnan Province Talent under grants 2012HB014, and the Natural Science Foundation of Yunnan Province under grant 2013FD014. This work is also supported by the Science Foundation of Yunnan educational department (grant 2012Z016).



\begin{thebibliography}{00}
\bibitem[Abramowski et al.(2013)]{Ab13}
Abramowski, A., Acero, F., Aharonian, F., et al. 2013, A\&A, 550, 4
\bibitem[Acciari et al.(2009)]{Ac09}
Acciari, V.A., Aliu, E., Arlen, T., et al. 2009, ApJ, 693, L104
\bibitem[Ackermann et al.(2012)]{Ac12}
Ackermann, M., Ajello, M., Allafort, A., et al. 2012, Sci, 338, 1190
\bibitem[Actis et al.(2011)]{Act11}
Actis, M., Agnetta, G., Aharonian, F., et al.  2011, Exp. Astron., 32, 193
\bibitem[Aharonian(2000)]{A00}
Aharonian F.A. 2000, New Astron., 5, 377
\bibitem[Aharonian et al.(2006a)]{A06a}
Aharonian, F., Akhperjanian, A. G., Bazer-Bachi, A. R., et al. 2006a, Natur, 440, 1081
\bibitem[Aharonian et al.(2006b)]{A06b}
Aharonian, F., Akhperjanian, A. G., Bazer-Bachi, A. R., et al. 2006b, Sci, 314, 1424
\bibitem[Aharonian et al.(2007)]{A07}
Aharonian, F., Akhperjanian, A.G., Barres de Almeida, U., et al.  2007, A\&A, 475, L9
\bibitem[Aharonian et al.(2010)]{A10}
Aharonian, F.A., Kelner, S.R., \& Prosekin, A. 2010, Phy. Rev. D, 82, 043002
\bibitem[Aharonian et al.(2013)]{Ah13}
Aharonian, F., Essey, W., Kusenko, A., \& Prosekin, A. 2013, Phy. Rev. D, 87, 063002
\bibitem[Aliu et al.(2014)]{Al14}
Aliu, E., Archambault, S., Arlen, T., et al.  2014, ApJ, 782, 13
\bibitem[Atoyan \& Dermer(2001)]{At01}
Atoyan, A., \& Dermer, C.D. 2001, Physical Review Letters, 87, 221102
\bibitem[Berezinsky et al.(2006)]{Be06}
Berezinsky, V., Gazizov, A.Z., \& Grigorieva, S.I. 2006, Phy. Rev. D, 74, 043005
\bibitem[Bloom \& Marscher(1996)]{Bl96}
Bloom S.D., \& Marscher A.P. 1996, ApJ, 461, 657
\bibitem[Blumenthal \& Gould(1970)]{Blu70}
Blumenthal, G.R. 1970, Phys. Rev. D, 1, 1596
\bibitem[B$\rm \ddot{o}$ttcher \& Dermer(1998)]{B98}
B$\rm \ddot{o}$ttcher, M.,\& Dermer, C.D. 1998, ApJ, 501, L51
\bibitem[Cao(2010)]{Cao10}
Cao, Z. 2010, Chin. Phys. C, 34, 249
\bibitem[Cerruti et al.(2015)]{Ce15}
Cerruti, M., Zech, A., Boisson, C., \& Inoue, S. 2015, MNRAS, 448, 910
\bibitem[Chiang \& Bottcher(2002)]{Ch02}
Chiang, J., \& B$\rm \ddot{o}$ttcher, M. 2002, ApJ, 564, 92
\bibitem[Chodorowski et al.(1992)]{Cho92}
Chodorowski, M.J., Zdziarski, A.A., \& Sikora, M. 1992, ApJ, 400, 181
\bibitem[Costamante et al.(2001)]{Co01}
Costamante, L., et al. 2001, A\&A, 371, 512
\bibitem[Costamante et al.(2008)]{Co08}
Costamante, L., Aharonian, F., Buhler, R., et al. 2008, in AIP Conf. Proc. 1085, High Energy Gamma-ray Astronomy, ed. F. A. Aharonian, W. Hofmann, \& E. Rieger (Melville, NY: AIP), 644
\bibitem[Costamante(2013)]{Co13}
Costamante, L. 2013, IJMPD, 22, 30025
\bibitem[Cui et al.(2014)]{Cu14}
Cui, S.W., Liu, Y., Liu, Y.J., et al. 2014, Astrop. Phys., 54, 86
\bibitem[Dai et al.(2002)]{Da02}
Dai, Z.G., Zhang, B., Gou, L.J., et al. 2002, ApJ, 580, L7
\bibitem[de Angelis et al.(2007)]{deA07}
de Angelis, A., Roncadelli, M., Mansutti, O. 2007, Phy. Rev. D, 76, 121301
\bibitem[Dermer et al.(1992)]{De92}
Dermer, C.D., Schlickeiser, R., \& Mastichiadis, A. 1992, A\&A, 256, L27
\bibitem[Dermer \& Schlickeiser(1993)]{DeS93}
Dermer, C.D., \& Schlickeiser, R. 1993, ApJ, 416, 458
\bibitem[Dermer et al.(2009)]{De09}
Dermer, C.D., Razzapue, S., Finke, J.D., \& Atoyan, A. 2009, New J. Phys., 11, 065016
\bibitem[Dermer \& Menon(2009)]{DM09}
Dermer, C.D. \& Menon, G. 2009, High Energy Radiation from Black Holes:Gamma Rays, Cosmic Rays, and Neutrinos, ed. Dermer, C. D. \& Menon, G.
\bibitem[Dimitrakoudis et al.(2012)]{Di12}
Dimitrakoudis, S., Mastichiadis, A., Protheroe, R.J., \& Reimer, A. 2012, A \& A, 546, 120
\bibitem[Dwek \& Krennrich(2005)]{Dw05}
Dwek, E., \& Krennrich, F. 2005, ApJ, 618, 657
\bibitem[Dwek \& Krennrich(2013)]{Dw13}
Dwek, E., Krennrich, F. 2013, APh, 43, 112
\bibitem[Essey \& Kusenko(2010)]{EK10}
Essey, W., \& Kusenko, A. 2010, Astropart. Phys., 33, 81
\bibitem[Essey et al.(2010)]{E10}
Essey, W., Kalashev, O.E., Kusenko, A., \& Beacom, J.F. 2010, PhRvL, 104, 141102
\bibitem[Essey et al.(2011a)]{E11a}
Essey, W., Ando, S., Kusenko, A. 2011a, Astropart. Phys., 35, 135
\bibitem[Essey et al.(2011b)]{E11b}
Essey, W., Kalashev, O.E., Kusenko, A., \& Beacom, J.F. 2011, ApJ, 731, 51
\bibitem[Fan et al.(2004)]{Fa04}
Fan, Y.Z., Dai, Z.G., \& Wei, D.M. 2004, A\&A, 415, 483
\bibitem[Finke \& Razzaque(2009)]{F09}
Finke, J.D., \& Razzaque, S. 2009, ApJ, 698, 1761
\bibitem[Franceschini(2008)]{Fr08}
Franceschini, A., Rodighiero, G., \& Vaccari, M. 2008, A\&A, 487, 837
\bibitem[Hampf(2011)]{H11}
Hampf, D., Tluczykont, M., \& Horns, D. 2011, Proc. of TEXAS 2010 conference in Heidelberg
\bibitem[Inoue et al.(2013)]{I13}
Inoue, Y., Kalashev, O.E., \& Kusenko, A. 2013, APh., 54, 118
\bibitem[Ghisellini \& Madau(1996)]{Gh96}
Ghisellini, G., \& Madau, P. 1996, MNRAS, 280, 67
\bibitem[Ghisellini et al.(2014)]{Gh14}
Ghisellini, G., Tavecchio, F., Maraschi, L., et al. 2014, Nature, 515, 376
\bibitem[Giommi et al.(1995)]{Gi95}
Giommi, P., Ansari, S.G., Micol, A. 1995, A\&AS, 109, 267
\bibitem[Gould \& Schreder(1966)]{GS66}
Gould, R., \& Schreder, G. 1966, Phys. Rev. Lett., 16, 252
\bibitem[Greisen(1966)]{Gr66}
Greisen, K. 1966, PhRvL, 16, 748
\bibitem[Kalashev et al.(2013)]{Ka13}
Kalashev, O.E., Kusenko, A., \& Essey, W. 2013, PhRvL, 111, 041103
\bibitem[Katarzynski et al.(2001)]{Kat01}
Katarzynski, K., Sol, H., \& Kus, A. 2001, A\&A, 367, 809
\bibitem[Kifune(1999)]{Ki99}
Kifune, T. 1999, ApJ, 518, L21
\bibitem[Kneiske et al.(2004)]{Kn04}
Kneiske, T.M., Bretz, T., Mannheim, K., \& Hartmann, D.H. 2004, A\&A, 413, 807
\bibitem[Konopelko et al.(2003)]{Ko03}
Konopelko, A., Mastichiadis, A., Kirk, J., de Jager, O.C., \& Stecker, F.W. 2003, ApJ, 597, 851
\bibitem[Lefa et al.(2011)]{Le11}
Lefa, E., Rieger, F.M., \& Aharonian, F. 2011, ApJ, 740, L64
\bibitem[Mannheim \& Biermann(1992)]{MB92}
Mannheim, K., \& Biermann, P.L. 1992, A\&A, 253, L21
\bibitem[Mannheim(1993)]{Ma93}
Mannheim, K., 1993, A\&A, 269, 67
\bibitem[Maraschi et al.(1992)]{Ma92}
Maraschi, L., Ghisellini, G., \& Celotti, A. 1992, ApJ, 397, L5
\bibitem[Mastichiadis \& Kirk(1997)]{MK97}
Mastichiadis, A., \& Kirk, J.G. 1997, A\&A, 320, 19
\bibitem[Mastichiadis \& Kirk(1995)]{MK95}
Mastichiadis, A., \& Kirk, J.G. 1995, A\&A, 295, 613
\bibitem[Mastichiadis et al.(2005)]{Ma05}
Mastichiadis, A., Protheroe, R.J., \& Kirk, J.G. 2005, A\&A, 433, 765
\bibitem[Maximon(1968)]{M68}
Maximon, L.C. 1968, J. Res. Natl. Bur. Std., 72, 79
\bibitem[Mazin \& Raue(2007)]{Ma07}
Mazin, D., \& Raue, M. 2007, A\&A, 471, 439
\bibitem[Murase(2009)]{Mu09}
Murase, K., \& Takami, H. 2009, ApJ, 690, L14
\bibitem[Murase(2012)]{Mu12}
Murase, K. 2012, ApJ, 745, L16
\bibitem[Murase et al.(2012)]{Mu12}
Murase, K., Dermer, C.D., Takami, H., \& Migliori, G. 2012, ApJ, 749, 63
\bibitem[M$\ddot{\rm u}$cke \& Protheroe(2001)]{MP01}
M$\ddot{\rm u}$cke A., \& Protheroe, R. 2001, Astropart. Phys., 15, 121
\bibitem[M$\ddot{\rm u}$cke et al.(2003)]{M03}
M$\ddot{\rm u}$cke, A., Protheroe R.J., Engel R., Rachen J.P., \& Stanev T. 2003, Astropart. Phys., 18, 593
\bibitem[Neronov(2012)]{Ne12}
Neronov, A., Semikoz, D., \& Taylor, A.M. 2012, A\&A, 541, 31
\bibitem[Nikishov(1962)]{N62}
Nikishov, A. 1962, J. Eep. Theor. Phys. Lett., 14, 393
\bibitem[Persic(2014)]{Per14}
Persic, M., \& Rephaeli, Y. 2014, A\&A, 567, 101
\bibitem[Petropoulou(2014)]{P14}
Petropoulou, M. 2014, MNRAS, 442, 3026
\bibitem[Petropoulou \& Mastichiadis(2015)]{PM15}
Petropoulou, M., \& Mastichiadis, A. 2015, MNRAS, 447, 36
\bibitem[Phol \& Schlickeiser(2000)]{PS00}
Pohl, M., \& Schlickeiser, R. 2000, A\&A, 354, 395
\bibitem[Protheroe \& Meyer(2000)]{PM00}
Protheroe, R.J., \& Meyer, H. 2000, Physics Letters B, 493, 1
\bibitem[Raiteri et al.(2012)]{R12}
Raiteri, C.M., et al. 2012, A\&A, 545, 48
\bibitem[Romero \& Vila(2008)]{R08}
Romero, G.E., \& Vila, G.S. 2008, A\&A, 485, 623
\bibitem[Sanchez et al.(2013)]{S13}
Sanchez, D.A., Fegan, S., \& Giebels, B. 2013, A\&A, 554, 75
\bibitem[Sanchez-Conde et al.(2009)]{S09}
Sanchez-Conde, M.A., Paneque, D., Bloom, E., Prada, F., Dominguez, A. 2009, Phy. Rev. D, 79, 123511
\bibitem[Sandoval et al.(2009)]{San09}
Sandoval, A., Alfaro, R., Belmont, E., et al. 2009, Fermi Symposium, eConf Proceedings C091122
\bibitem[Schlickeiser(2002)]{Sch02}
Schlickeiser, R., 2002, Cosmic Ray Astrophysics (Berlin: Springer), p.472
\bibitem[Sikora et al.(1994)]{S94}
Sikora, M., Begelman, M.C., \& Rees, M.J. 1994, ApJ, 421, 153
\bibitem[Simet et al.(2008)]{S08}
Simet, M., Hooper, D., Serpico, P.D. 2008, Phy. Rev. D, 77, 063001
\bibitem[Stecker et al.(1992)]{S92}
Stecker, F.W., de Jager, O.C., \& Salamon, M.H. 1992, ApJ, 390, L49
\bibitem[Stecker et al.(2006)]{S06}
Stecker, F.W., Malkan, M.A., \& Scully, S.T. 2006, ApJ, 648, 774
\bibitem[Sobolewska et al.(2014)]{S14}
Sobolewska, M.A., Siemiginowska A., Kelly, B.C., \& Nalewajko, K. 2014, ApJ, 786, 143
\bibitem[Takami et al. (2013)]{T13}
Takami, H., Murase, K., \& Dermer, C.D. 2013, ApJ, 771, L32
\bibitem[Tavecchio et al. (2010)]{Ta10}
Tavecchio, F., Ghisellini, G., Foschini, L., Bonnoli, G., Ghirlanda, G., Coppi, P. 2010, MNRAS, 406, L70
\bibitem[Urry \& Padovani(1995)]{U95}
Urry, C.M., \& Padovani, P. 1995, PASP, 107, 803
\bibitem[Urry(1998)]{U98}
Urry, C.M., Adv. Space Res. 1998, 21, 89
\bibitem[Venters(2010)]{V10}
Venters, T.M. 2010, ApJ, 710, 1530
\bibitem[Vovk et al.(2012)]{V12}
Vovk, I., Taylor, A.M., Semikoz, D., \& Neronov, A. 2012, ApJ, 747, L14
\bibitem[Wagner (2008)]{Wa08}
Wagner, R.M. 2008, MNRAS, 385, 119
\bibitem[Woo et al.(2005)]{W05}
Woo, J.H., Urry, C.M., van der Marel, R.P., Lira, P., \& Maza, J. 2005, ApJ, 631, 762
\bibitem[Wu et al.(2007)]{Wu07}
Wu, Z.Z., Jiang, D.R., Gu, M.F., \& Liu, Y. 2007, A\&A, 466, 63
\bibitem[Yang et al.(2008)]{Y08}
Yang, C.Y., Fang, J., Lin, G.F., Zhang, L. 2008, ApJ, 682, 767
\bibitem[Zatsepin \& Kuzmin(1966)]{Z66}
Zatsepin, G.T., \& Kuzmin, V.A. 1966, JETPL, 4, 78
\bibitem[Zheng \& Zhang(2011)]{Z11}
Zheng, Y.G., \& Zhang, L. 2011, ApJ, 728, 105
\bibitem[Zheng \& Kang(2013)]{Z13}
Zheng, Y.G., \& Kang, T. 2013, ApJ, 764, 113

\end{thebibliography}
\end{document}